\documentclass[twocolumn,showpacs,preprintnumbers,amsmath,amssymb]{revtex4}

\usepackage{graphicx}
\usepackage{dcolumn}
\usepackage{bm}

\begin{document}

\preprint{APS/123-QED}

\title{Optical properties of $MgCNi_{3}$ in the normal state}

\author{P. Zheng}
\author{J. L. Luo }
\author{G. T. Liu }
\author{Y. L. Zhang }
\author{R. C. Yu }
\author{C. Q. Jin }
\author{N. L. Wang }
\altaffiliation[Email: ]{nlwang@aphy.iphy.ac.cn}
\affiliation{Beijing National Laboratory for Condensed Matter
Physics, Institute of Physics, Chinese Academy of Sciences,
Beijing 100080, P. R. China}%

\author{M. Q. Tan}
\affiliation{Department of Physics, Zhejiang University, Hangzhou 310027, P. R. China}%


\begin{abstract}
We present the optical reflectance and conductivity spectra for
non-oxide antiperovskite superconductor $MgCNi_{3}$ at different
temperatures. The reflectance drops gradually over a large energy
scale up to 33,000 cm$^{-1}$, with the presence of several
wiggles. The reflectance has slight temperature dependence at low
frequency but becomes temperature independent at high frequency.
The optical conductivity shows a Drude response at low frequencies
and four broad absorption features in the frequency range from 600
$cm^{-1}$ to 33,000 $cm^{-1}$. We illustrate that those features
can be well understood from the intra- and interband transitions
between different components of Ni 3d bands which are hybridized
with C 2p bands. There is a good agreement between our
experimental data and the first-principle band structure
calculations.
\end{abstract}


\maketitle
The discovery of non-oxide superconductor $MgCNi_{3}$
has attracted much attention due to the coexistence of
ferromagnetic element $Ni$ and superconductivity\cite{He}.
$MgCNi_{3}$ has the cubic antiperovskite lattice structure with
space group $Pm\bar{3}m$, in which eight $Mg$ atoms occupy the
cubic corners, while six \textit{Ni} atoms are at the face-center
positions and one \textit{C} atom is inserted into the body-center
position. The lattice parameter $a$ is  $3.81 \AA$ \cite{He}. Hall
effect \cite{Li1}, thermopower and thermal conductivity \cite{Li2}
measurements indicate electron-type conducting carriers in
$MgCNi_{3}$. A number of experiments including $C^{13}$ NMR
measurement below $T_{c}$ \cite{Singer}, the determination of
upper critical field $H_{c2}(0)$ from the resistance $\rho(T)$
under different magnetic fields \cite{Li1}, and the specific heat
measurements\cite{Lin} show that $MgCNi_{3}$ is a conventional
superconductor. But band calculations indicate that the $Ni$ $3d$
electrons dominate the density of the states (DOS) near the Fermi
level $E_{F}$\cite {Szajek, Dugdal, Singh, Kim}. Then, the $Ni$
$3d$ electrons are usually considered to take part in the
superconducting pairing in $MgCNi_{3}$. In such a case,  an s-wave
pairing is unexpected because of the existence of localized moment
of Ni atom which will break the s-wave superconducting pairing.

The relationship between the electronic states of $Ni$ and the
superconductivity in $MgCNi_{3}$ was intensively studied. All of
the known substitutions at Ni-site are found to decrease the
superconducting temperature and/or bulk superconducting volume
\cite{Das,Hayward, Kumary}. Normal state $^{13}C$ NMR measurement
implies the existence of spin fluctuation above $20$
$K$\cite{Singer}. The calculated Stoner exchange parameter $S=
0.43$ $\sim$ $0.64$\cite{Dugdal, Shim} and the Stoner
renormalization is 5\cite{Singh}. The value is at the high edge of
paramagnetic region and close to the ferromagnetic region which
may lead to ferromagnetic spin fluctuation. Apparently, $Ni$ is
related with the superconductivity but its electronic and magnetic
properties are different from those in pure f.c.c $Ni$.
Considering that $MgCNi_{3}$ has the isomorphic lattice as the
pure $Ni$, the effect of $Mg$, $C$ atoms should be important to
change the electronic and the magnetic properties of $Ni$ in
$MgCNi_{3}$. Dugdale and Jarlborg\cite{Dugdal} studied the effect
of $Mg$, $C$ on $MgCNi_{3}$ in comparison with the $f.c.c$ $Ni$.
Their calculation suggests that for $MgCNi_{3}$, both the
face-center substitutions of $f.c.c$ $Ni$ by $Mg$ atom and the
occupation of the C-site at the body-center position will make the
Fermi level $E_{F}$ of $MgCNi_{3}$ to move away from the large DOS
peak and make $N(E_{F})$ smaller than that of $f.c.c$ $Ni$,
leading to a decrease of the Stoner factor and a disappearance of
the magnetic order. Other theoretical studies show that\cite{Shim,
Singh,Walte}, the correlation of $Ni$ $3d$ electrons in
$MgCNi_{3}$ is not very important like in f.c.c $Ni$ due to strong
hybridization between $Ni$ $3d$ and $C$ $2p$ states.

The optical conductivity has been calculated based on the first
principle band structure study\cite{Okoye,Tan}. According to the
local orbital symmetry of $Ni$, the $Ni$ $3d$ states should be
decomposed into $xy$, $yz+zx$, $x^{2}-y^{2}$ and $3z^{2}-r^{2}$
components. They contribute to the region from the +1 to -4 eV.
The C p orbitals are hybridized with Ni d, and are located below
most of the the Ni d states. The partial DOS calculations further
indicate that the states very close to $E_{F}$ are derived by $Ni$
$3d_{yz+zx}$ and $3d_{3z^{2}-r^{2}}$ orbitals, which also dominate
the chemical bonding between $C$ and ligand $Ni$ atoms. The
contributions by Ni 3d$_{xy}$ and $x^{2}-y^{2}$ are somewhat away
from $E_{F}$. The calculated conductivity spectrum at low energy
is dominated by intra- and interband transitions among those Ni 3d
states. The spectrum appears quite different from that of pure
f.c.c $Ni$ which has only two peaks at $0.3$ $eV$ and $1.4$ $eV$
below $4$ $eV$ at $300$ $K$\cite{Ehrenreich}. But up to now, as
far as we know, there is no experimental study on the optical
properties of $MgCNi_{3}$ compound. In this work, we report our
optical reflectance investigation on the $MgCNi_{3}$
superconductor from $50$ $cm^{-1}$ to $46,000$ $cm^{-1}$ at
different temperatures. Our study shows that $Ni$ $3d$ bands
really dominate the electronic structure near Fermi level, but the
$0.3$ $eV$ and $1.4$ $eV$ peaks which is related to the
ferromagnetic property of pure f.c.c $Ni$ disappear in
$MgCNi_{3}$. A narrow Drude response exists in low frequency
region. There are four interband transitions below $4$ $eV$. The
origin of the intra- and interband transitions is discussed.

\begin{figure}
\includegraphics[width=0.9\linewidth]{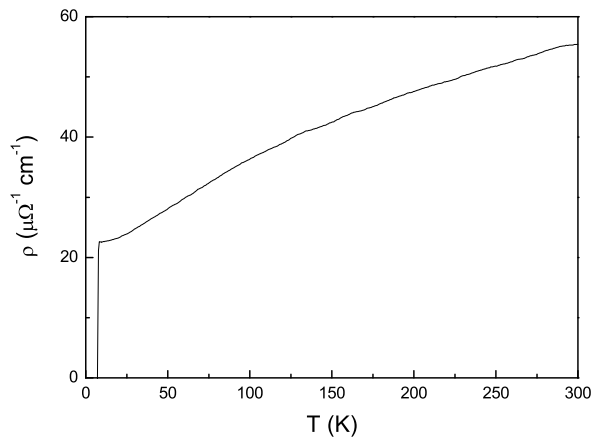}
\caption{\label{fig:epsart} The temperature dependence of the
resistivity of $MgCNi_{3}$. }
\end{figure}

Because there is no available single crystal of $MgCNi_{3}$
compound, we performed our optical study on polycrystalline
$MgCNi_{3}$ sample prepared by conventional solid state reaction
followed by post high pressure treatment. A single phase
polycrystalline $MgCNi_{3}$ compound was first synthesized using
solid state reaction as described in ref.\cite{Jin}. Then, the
compound was further treated at 800$^0$C for 5 minutes under
pressure of 5 GPa. The obtained sample is extremely dense and
checked again to be pure phase by X-ray diffraction. A very shinny
and metallic bright surface was obtained after fine polishing.
Since the material has a cubic structure, optical constants of
$MgCNi_{3}$ should be isotropic, therefore we can determine its
optical constants from the reflectance measurement on such
polycrystalline sample. The frequency-dependent reflectance
$R(\omega)$ was measured from $50$ $cm^{-1}$ to $46,000$ $cm^{-1}$
at different temperatures. The measurements below $25,000$
$cm^{-1}$ were performed on a Bruker 66v/s spectrometer. A grating
type spectrometer was used for the measurement above 20,000
$cm^{-1}$. Good agreement is seen in the overlapped frequency
region. An \textit{in-situ} gold (below 15,000 cm$^{-1}$) and
aluminum (above 15,000 cm$^{-1}$) overcoating technique was used
for the experiment, then the data were corrected for the absolute
reflectivity of gold and aluminum. We use Hagen-Rubens relation
for the low frequency extrapolation, and a constant extrapolation
to 100,000 cm$^{-1}$ followed by a well-known function of
$\omega^{-4}$ in the higher-energy side.

Fig. 1 displays the temperature dependence of the resistivity
measured by a standard four-probes method. It shows metallic
behavior in the normal state. Its superconducting temperature $Tc$
is 7.8 $K$ and $\Delta$$T|_{Tc}$ = 1.7 $K$, showing a high quality
of our sample. Fig. 2 shows the frequency dependent reflectance
$R(\omega)$ at $300$ $K$ from $50$ $cm^{-1}$ to $46,000$
$cm^{-1}$. The inset shows the $R(\omega)$ spectra at $300$ $K$,
$180$ $K$ and $10$ $K$ from $50$ $cm^{-1}$ to $8,000$ $cm^{-1}$.
$R(\omega)$ at $300$ $K$, $180$ $K$ and $10$ $K$ cross near
$3,000$ $cm^{-1}$. The reflectivity of $MgCNi_{3}$ monotonically
decreases with the frequency up to 33,000 $cm^{-1}$, showing
typical over-damped characteristic. Several wiggles could be
observed in this energy region. Such a shape is roughly similar to
the frequency dependence of reflectivity of pure f.c.c.
$Ni$\cite{Ehrenreich}. As reported, the gradual drop of the
reflectivity in this range appears to be a characteristic behavior
of most metals like $Fe$, $Co$, and Pd in which the $d$ bands play
a prominent role\cite{Ehrenreich}. The similarity of optical
spectra between $MgCNi_{3}$ and pure $Ni$ metal implies that $Ni$
$3d$ bands dominate the electronic structure of $MgCNi_{3}$ below
$33,000 cm^{-1}$ and is consistent with the theoretical analysis.
However, differences between the two materials are also remarkable
especially in high frequency region: (1) In the reflectance
spectrum of $MgCNi_{3}$, there are two more broad peaks than pure
$Ni$. (2) The central positions of the other two broad peaks have
moved, as we should illuminate below.

\begin{figure}
\includegraphics[width=0.9\linewidth]{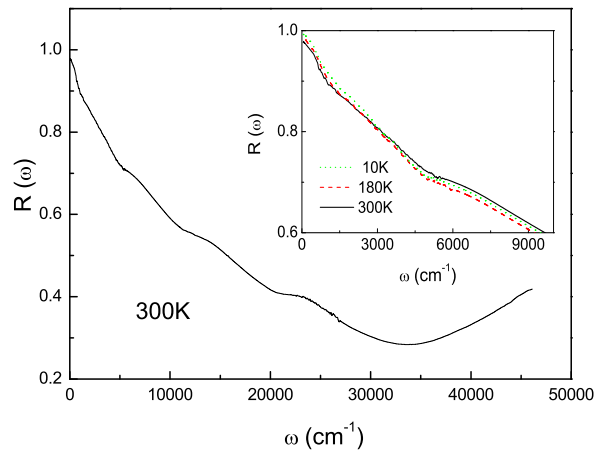}
\caption{\label{fig:epsart}(Color online) The frequency dependence
of reflectivity $R(w)$ at $300$ $K$ in the frequency region from
$50$ $cm^{-1}$ to $45,000$ $cm^{-1}$. The inset shows the spectra
of $R(w)$ $300$ $K$,$180$ $K$ and $10$ $K$ from $50$ $cm^{-1}$ to
$8,000$ $cm^{-1}$, respectively.}
\end{figure}

Fig. 3 shows the real part of the optical conductivity
$\sigma(\omega)=\sigma_{1}(\omega)+i\sigma_{2}(\omega)$ below
$40,000$ $cm^{-1}$ at $300$ $K$, $180$ $K$ and $10$ $K$. The
conductivity at low-frequency limit $\sigma_1(\omega)$ $\approx$
$20,000$ $\Omega^{-1}cm^{-1}$, which is close to the value at
$300$ $K$ deduced from Fig. 1. A Drude response exists in the low
frequency region below $600$ $cm^{-1}$. In the frequency region
from $600$ $cm^{-1}$ to $33,000$ $cm^{-1}$, there are four broad
peaks centering at $1,300$ $cm^{-1}$, $5,700$ $cm^{-1}$, $12,700$
$cm^{-1}$ and $23,300$ $cm^{-1}$(i.e. $0.16$ $eV$, $0.7$ $eV$,
$1.6$ $eV$ and $2.84$ $eV$) respectively. The four peaks keep
present down to $10$ $K$. The three peaks at high frequency are
hardly temperature-dependent, while the low frequency peak and the
Drude response show slight temperature dependence: The height of
the broad peak at $1,300$ $cm^{-1}$ decreases when temperature
decreases to $10$ $K$ but the peak position almost does not move.

\begin{figure}
\includegraphics[width=0.9\linewidth]{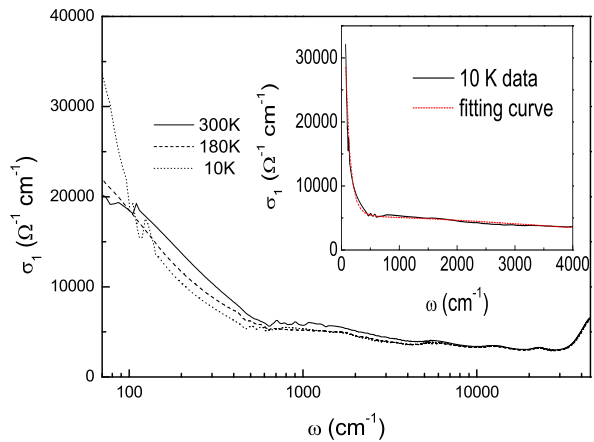}
\caption{\label{fig:epsart}(Color online) Frequency dependence of
the optical conductivity $\sigma_{1}(\omega)$ at different
temperatures. The solid line is for $300$ $K$, the dashed line is
for $180$ $K$ and the dotted line is for $10$ $K$. The inset shows
the fitting result at $10$ $K$, in which the solid line is the
experimental data and the dashed line is the fitting line.}
\end{figure}

For f.c.c $Ni$, besides a Drude response, there are only two broad
peaks centering at about $0.3$ $eV$ and $1.4$ $eV$ in the
frequency below $4.0$ $eV$ at 300 K\cite{Ehrenreich}. In the
ferromagnetic state below the  Curie temperature, the exchange
interaction lowers the energy of the spin-up electrons and raises
the energy of the spin-down electrons of the d bands. This leads
to two subsets of electrons in the region of the Brillouin zone
surrounding $L$ point. The $0.3$ $eV$ peak in pure f.c.c $Ni$ is
due to the interband transition between the spin-down electrons of
the 3d band and the 4s band at the Fermi surface. The $1.4$ $eV$
is related to the interband transition from spin-up band of 3d
electrons to the 4s band at the Fermi level. For $MgCNi_{3}$,
there are four broad peaks centering at about $ 0.16$ $eV$,
$0.71$, $1.59$ and $2.84$ $eV$ respectively. We notice that both
the $0.3$ $eV$ and $1.4$ $eV$ peaks in pure f.c.c $Ni$ disappear
in $MgCNi_{3}$. Instead, the number of peaks below $4.0$ $eV$ has
increased to four. Considering the small Stoner
factor\cite{Dugdal,Shim} and the dc susceptibility\cite{Das}, the
four peaks must have different origin as those observed in the
optical conductivity spectra of pure f.c.c $Ni$.

According to the theoretical analysis for
$MgCNi_{3}$\cite{Singh,Tan}, $Ni$ $3d$ bands decompose into $xy$,
$yz+zx$, $x^{2}-y^{2}$ and $3z^{2}-r^{2}$ four components in a
local tetragonal crystal field\cite{Singh,Tan}. The energy range
of band $Ni$ $3d_{xy}$ is $-4.0$ $eV$ $\sim$ $-0.5$ $eV$.  Its
main DOS peaks are at about $-1.0$ $eV$, $-1.3$ $eV$ with a weaker
peak at $-2.7$ $eV$. The band $Ni$ $3d_{yz+zx}$ extends in energy
region from $-4$ $eV$ to $+1$ $eV$ with two high DOS peaks at
about $-2.5$ $eV$, $-0.08$ $eV$ respectively. $Ni$
$3d_{x^{2}-y^{2}}$ has some DOS from $-4$ $eV$ and 0 $eV$ with the
highest peak at about $-1.7$ $eV$. Band $Ni$ $3d_{3z^{2}-r^{2}}$
locates in the range of $-4$ $eV$ to $0$ $eV$. It also contributes
to the states very close to the Fermi level. The $C$ $2p$ bands
hybridize with $Ni$ $3d$ bands. Its main DOS is in energy range of
$-7.0$ $eV$ to $-4.0$ $eV$, although $C$ $2p$ bands offer a few
DOS near $E_{F}$ from about $-0.1$ $eV$ to $+1$ $eV$ as well. The
three peaks in the energy range of $0.5$ $eV$ to $4$ $eV$ observed
in Fig. 2 could be attributed to the interband transitions from
the occupied $Ni$ $3d$ states to unoccupied part of Ni 3d state
(the major contribution is $Ni$ $3d_{yz+zx}$ bands) which
hybridized strongly with the $C$ $2p$ states\cite{Note}. The peak
of $2.84$ $eV$ can be assigned mainly to the interband transition
from the occupied $Ni$ $3d_{yz+zx}$ bands to another unfilled part
of d band with mixture of unoccupied $C$ $2p$ band, while the
$-2.7$ $eV$ DOS peak of $Ni$ $3d_{xy}$ may offer small spectra
weight too. The $1.59$ $eV$ peak may be dominated by the interband
transition from the occupied $-1.3$ $eV$ peak of $Ni$ $3d_{xy}$
bands to the unoccupied $Ni$ $3d_{yz+zx}$/$C$ $2p$ mixing states,
while the peak of $0.71$ $eV$ may be mainly due to the interband
transition from $Ni$ $3d_{3z^{2}-r^{2}}$ band to the unoccupied
$Ni$ $3d_{yz+zx}$/$C$ $2p$ mixing states. The three peaks of
$0.71$ $eV$, $1.59$ $eV$, $2.84$ $eV$ in the experiment is quite
consistent with previous theoretical calculation based on the
first principle band structure study\cite{Tan}. The calculations
indicate three peaks at about $0.73$ $eV$, $1.8$ $eV$, and $2.9$
$eV$ in the real part of conductivity $\sigma_{1}(\omega)$ in the
region from $0.5$ $eV$ to $4$ $eV$.

The fourth broad peak in our optical conductivity spectra at about
$1,300$ $cm^{-1}$ has not been discussed in theoretical
study\cite{Tan}. As we describe above, this peak shows slight
temperature dependence. When temperature decreases, the weight of
the peak decreases but the peak position does not move. We find
that the sum of one Drude component and one Lorentz component
could fit the data at different temperatures below $4,000$
$cm^{-1}$:
\begin{eqnarray}
  \sigma_{1}(\omega)=\frac{\omega_{p}^{2}}{4\pi}\frac{\gamma}{(\omega^{2}+\gamma^{2})}+\frac{\omega_{p,1}^{2}}{4\pi}\frac{\omega^{2}\gamma_{1}}{(\omega^{2}-\omega_{0}^{2})^{2}+\gamma_{1}^{2}\omega^{2}}.
\end{eqnarray}
where $\omega_{p}$ is the characteristic plasma frequency of free
carriers, $\omega_{p,1}$ is the strength of the bounded carriers
associated with the Lorentz component, $\gamma$ is the inverse of
life time$(1/\tau)$ and $\gamma_{1}$ is the damping coefficient of
the Lorentz component. The inset of Fig. 3 shows the fitting
result at $10$ $K$. The fitting parameters are listed in table.I.

\begin{table}
\caption{\label{tab:table1}The fitting parameters of Drude and
Lorentz components for the conductivity spectra of $MgCNi_{3}$ at
$300 K$, $180$ $K$ and $10$ $K$.}
\begin{ruledtabular}
\begin{tabular}{cccccc}
$T(K)$ & $\omega_{p}$ & $\gamma$ & $\omega_{p,1}$ & $\gamma_{1}$ & $\omega_{0}$\\
300& 17000 & 210 & 41700 & 5700 & 1300\\
180& 16300 & 170 & 39300 & 5600 & 1300\\
 10& 16100 & 98  & 41200 & 6000 & 1300\\
\end{tabular}
\end{ruledtabular}
\end{table}

As we know, the Drude component is due to the intraband transition
of conducting carriers. It is mainly contributed by the electrons
of $Ni$ $3d_{yz+zx}$ and $Ni$ $3d_{3z^{2}-r^{2}}$ bands, which are
hybridized strongly with C 2p bands and cross the Fermi level.
Then, the major concern here is the assignment of the Lorentz
component. Because of the substantial Ni-C covalent interaction,
the electron-phonon coupling is expected to be strong in this
compound\cite{Singh,Tan}, which could lead to polaronic
characteristic of charge carriers. On this basis, one may link the
broad component to the photoionization effect of polarons.
However, the difficulty of this interpretation is that the
temperature dependence of the spectral weight of the component is
not consistent with the expected behavior of polarons. The peak
strength should increase with decreasing temperature for polaronic
absorption\cite{Yoon}, but in the present compound, the peak
spectral weight slightly decreases with decreasing temperature. An
alternative explanation is that this electronic band also
originates from an interband transition. Since the energy scale of
the transition, 1,300 cm$^{-1}$ (0.16 eV), is quite small, by
looking at the calculated band structure along different symmetry
lines in Brillouin zone\cite{Singh,Tan}, we found the transition
between the two bands close to the half way along the $\Gamma$-M
line possible. This is likely corresponding to the transition from
the states of $Ni$ $3d_{3z^{2}-r^{2}}$ band just below the Fermi
level to the unoccupied part of the $Ni$ $3d_{yz+zx}$ band
hybridized with C 2p band.

It deserves to remark that the plasma frequency $\omega_{p}$ of
the Drude term listed in table I is only about $17,000$ $cm^{-1}$,
which is much smaller than the calculated value\cite{Tan}. We
stress that this plasma frequency is different from that deduced
by integrating the optical conductivity spectrum
($\sigma_{1}(\omega)$) from zero frequency to a cutoff frequency
$\omega_{c}$, which usually is taken at the reflectance edge or
minimum position of $R(\omega)$. Obviously, in the later case, it
overestimates the plasma frequency because of the inclusion of
four interband excitations.

In summery, in this paper we report the experimental measurement
of the reflectance spectra $R(\omega)$ and the deduced frequency
dependence of the optical conductivity of $MgCNi_{3}$. Our
experimental data and analysis show that $Ni$ $3d$ states really
dominate the DOS below $4$ $eV$. The contributions of the free
carriers and interband transitions to the conductivity spectra are
discussed on the basis of band structure calculations. There is a
good agreement between our experimental data and the previous
theoretical calculations.

\begin{acknowledgments}
This work is supported by National Science Foundation of China and
the Knowledge Innovation Project of Chinese Academy of Sciences.
\end{acknowledgments}


\begin{thebibliography}{99}
\bibitem{Note} Generally speaking, the d-d transition is not allowed. However, in solids,
because of the hibridizations between the d bands and some other
bands from other atoms, the d-d interband transition is usually
observed in optical spectra.
\end{thebibliography}
\end{document}